\pdfoutput=1
\documentclass[12pt]{article}
\usepackage[]{graphicx}
\def\beq{\begin{equation}}
\def\eeq{\end{equation}}
\def\beqa{\begin{eqnarray}}
\def\eeqa{\end{eqnarray}}
\def\infint{\int_{-\infty}^\infty}
\begin{document}
\title{Micro-pixel accuracy centroid displacement estimation and detector calibration}
\author{Chengxing Zhai, Mike Shao, Renaud Goullioud, and Bijan Nemati\\
{\it Jet Propulsion Laboratory, California Institute of Technology, Pasadena, CA 91109}}
%\affil{Jet Propulsion Laboratory, California Institute of Technology, Pasadena, CA 91109.}
\maketitle
\abstract{
Precise centroid estimation plays a critical role in 
accurate astrometry using telescope images.
Conventional centroid estimation fits a template point spread function (PSF)
to the image data. Because the PSF is typically not known to high
accuracy due to wavefront aberrations and uncertainties in optical system,
a simple Gaussian function is commonly used.
PSF knowledge error leads to systematic errors in the conventional
centroid estimation. In this paper, we present an accurate centroid
estimation algorithm by reconstructing the PSF from well sampled
(above Nyquist frequency) pixelated images.
In the limit of an ideal focal plane array whose pixels have identical response
function (no inter-pixel variation), this method can estimate centroid
displacement between two 32$\times$32 images to sub-micropixel accuracy.
Inter-pixel response variations exist in real detectors, {\it e.g.}~CCDs,
which we can calibrate by measuring the pixel response of each pixel in
Fourier space. The Fourier transforms of the inter-pixel variations
of pixel response functions can be conveniently expressed in terms of
powers of spatial wave numbers using their Taylor series expansions.
Calibrating up to the third order terms in this expansion,
we show that our centroid displacement estimation is accurate
to a few micro-pixels using simulated data. This algorithm is applicable to
the new proposed mission concept Nearest Earth Astrometry Telescope 
(NEAT) to achieve mirco-arcsecond accuracy in relative astrometry for detecting
terrestrial exoplanets. This technology is also applicable to high
precision photometry missions.}

\section{Introduction}
Consider an $N{\times}N$ detector array with coordinate $(m,n)$
labeling the pixel in $m$-th row and $n$-th column.
Given the intensities $I_{mn}$ measured by pixels $(m,n), m,n=1,2,\cdots,N$,
a straight forward estimate of centroid is
\beq
  {\hat x}_c = {\sum_{m,n} x_{mn} I_{mn} \over \sum_{mn} I_{mn}} \,, \quad
  {\hat y}_c = {\sum_{m,n} y_{mn} I_{mn} \over \sum_{mn} I_{mn}} \,,
\label{basicCent}
\eeq
where $(x_{mn}, y_{mn})$ are the x and y coordinates of the center of
pixel $(m,n)$ and the summation is over all the pixels relevant to
the centroid estimation. Typically only a small array of pixels is used
for estimation because enlarging the size of array degrades
the signal to noise ratio (SNR) significantly.
The pixels at large distances from the center of the image detect very
little photons but are heavily weighted by their large coordinates.
Estimation~(\ref{basicCent}) suffers systematic errors from
using a relatively small array of pixels.
The point spread function (PSF) fitting algorithms \cite{centerFitting}
supersede this straight forward centroid estimation because it avoids
amplification of noise from multiplying large coordinates.
The main challenge for the PSF fitting approach is the knowledge of the PSF.
Computing PSF using a diffraction model requires both knowledge of the
optical system and wave front aberrations that both are usually hard to obtain.
In the past, Gaussian functions has been popularly used\cite{centerFitting}
However, in order to achieve a highly accurate centroid estimation,
a more precise PSF is needed.
Mighell has done work on PSF fitting using the digital images
by using 21-point damped sinc interpolation\cite{digitalPsf}.
In this paper, we work along the same line to reconstruct the PSF from
the pixelated images using the fact that the PSF is a bandwidth limited signal.
Theoretically, bandwidth limited signal can be reconstructed to
any accuracy as far as the sampling is above the Nyquist frequency
and the number of the sample is sufficiently large.
Using 32$\times$32 images, the truncation error causes
less than a micro-pixel error in centroid displacement estimation.

Due to the complex micro-structure of detectors and charge diffusion effect
between pixels, the pixel detection response varies over the
physical detection area of a pixel\cite{intraPixelVar}, which
is referred as {\it intra}-pixel variation.
The pixel counts is a convolution of the pixel response function
with the photon energy flux function. If each pixel has identical
response function, pixel counts still represents a bandwidth limited function
sampled at the pixel grid and thus the sampling theorem is still applicable
for reconstructing an effective PSF for the detector.
Past measurements found dominant intra-pixel detection variation
is indeed common to pixels \cite{prfDef1}. For micro-pixel centroid
estimation, we still need to take into account the
inter-pixel differences of the pixel response functions.
The inter-pixel response variations make the pixel counts no
longer represent sample values of a strictly bandwidth limited function.
To characterize inter-pixel variations, we use laser
metrology to measure the pixel responses in Fourier space.
It is convenient to parametrize the Fourier transforms of
the pixel response functions in terms of powers of spatial wave numbers
using their Taylor series expansions.
The leading order effect of inter-pixel variations
is the average pixel response or flat-field response,
which can be measured as response to a uniform E-field.
The first order correction is an effective geometric pixel location shift for
each pixel. The second and third order corrections are quadratic
and cubic polynomials of the spatial wave numbers. 
As we include more terms in the expansion, the model becomes more accurate
and the centroid estimation becomes more accurate.

In this paper, we present results based on simulations to demonstrate
the capability of calibrating the inter-pixel variations
for performing micro-pixel level centroid estimations by
including up to third order terms in the Taylor series expansion
of the Fourier transforms of the pixel response functions.
This algorithm can be utilized by the proposed mission NEAT\cite{NEAT}
to perform micro-arcsecond level relative astrometry with a one meter
telescope and thus detect terrestrial exo-planets in habitable zone.

In addition to precise astrometry, our pixel calibration technique is also
applicable to high precision photometry, which requires to characterize the
pixel response functions. With the calibration of pixel responses
and the reconstruction of PSF, the photometry is no longer sensitive to
pointing errors and detector response variations, which limit the
performance. Because we measure
the pixel responses in Fourier space, it is especially 
convenient for studying photometry when the images are not too under-sampled
because we only need to measure the pixel response to a modest spatial
frequency range to cover the bandwidth of the PSF.

\section{Model and Algorithm description}
\subsection{Pixel intensity model description}
To simplify the formulation, we assume that the image is
stable over a sampling period so that the temporal integration
by the detector is simply an overall factor of the duration of the
sampling period.

The image is represented by photo-electron counts recorded by
the pixels. A model describing the photo-electron counts
recorded by pixel in the $m$-th row and $n$-th column is expressed as
\beq
  I_{mn} (x_c, y_c) = \infint dx \infint dy I(x - x_c, y - y_c) Q_{mn}(x,y)
\label{pixelModel}
\eeq
where $Q_{mn}(x,y)$ is the pixel response function (PRF) of pixel $(m,n)$
to a point illumination at $(x,y)$ in the detector plane\cite{prfDef1,prfDef2},
$I(x,y)$ is an input intensity function, and $(x_c,y_c)$ is the location of
the centroid of the image. For a point source, which we will
consider exclusively, $I(x,y)$ is the point spread function (PSF) related 
to the focal plane E-field via
\beq
 I(x,y) = \left | E(x,y) \right |^2 \,. 
\eeq
By Fourier optics, $E(x,y)$ is related to the E-field at the input pupil
of the telescope $E_i(x,y)$ via a Fourier transform
\beq
  E(x,y) = {\cal N}\infint dx^\prime \infint dy^\prime
  P(x^\prime,y^\prime)E_{\rm i}(x^\prime,y^\prime)
  \exp \left \{ {i2\pi \over \lambda f} (x x^\prime + y y^\prime) \right \}
\label{pupilFT}
\eeq
where ${\cal N}$ is a normalization factor,
$\lambda$ is the wavelength of the light, $f$ is the focal length
of the telescope, and $P(x,y)$ is the aperture function, whose value is 1 inside
the aperture and 0 outside the aperture.
We will focus on monochromatic light case and briefly discuss about the
polychromatic case in section~\ref{sec:polychrom}. Polychromatic case
requires one extra integral over the wave number,
which makes the formulation slightly complicated. The main steps
of derivation remain the same.

Performing a change of variable
$(k_x, k_y) = 2\pi / \left (\lambda f\right ) \left (x^\prime, y^\prime \right )$,
expression~(\ref{pupilFT}) becomes a two dimensional Fourier integral for $E(x,y)$ as
\beq
  E(x,y) = {\cal N} \left ({\lambda f \over 2 \pi} \right )^2 \infint dk_x \infint dk_y
  P\left({\lambda f \over 2 \pi} k_x, {\lambda f \over 2 \pi} k_y \right ) 
  E_{\rm i} \left({\lambda f \over 2 \pi} k_x, {\lambda f \over 2 \pi} k_y \right ) e^{ i(k_x x + k_y y)}
\label{pupilFT2}
\eeq
where $(k_x, k_y)$ represents the spatial frequency.
Because $P(x,y)$ vanishes outside the aperture of the telescope, $E(x,y)$ is a two dimensional
bandwidth limited signal. By the theorem of convolution, the Fourier transform of $I(x,y)$ is
the Fourier transforms of $E(x,y)$ convolved with its complex conjugate.
Therefore, $I(x,y)$ is also a bandwidth limited function with bandwidth being
twice of that of $E(x,y)$ from the process of convolution.
Let $D$ be the diameter of the aperture of the telescope, the bandwidth of
$I(x,y)$ is then limited by $|k_x| < 2\pi D/(\lambda f), |k_y| < 2\pi D/(\lambda f)$.%
\footnote{In fact, it can be shown that the non-zero Fourier frequencies
satisfies $k_x^2 + k_y^2 \le \left (2\pi D / (\lambda f)\right)^2$.}
Assuming the pixel size is smaller than the Nyquist
sampling spacing $f \lambda /(2D)$, by sampling theorem,
the intensity function $I(x,y)$ can be precisely reconstructed from
a pixelated image given an infinitely large detection array.
In practice, this is complicated by two things.
First of all, we can not have infinite samples. Therefore,
we have to truncate the reconstruction process at some finite size,
which introduces truncation error. We found that for 32$\times$32 array size
(including about the 7th Airy ring)
the truncation errors are small enough for micro-pixel centroid estimation.
Secondly, the pixel intensity model involves pixel response functions
$Q_{mn}(x,y)$, which depend on pixels.
If $Q_{mn}(x,y)$ does not depend on pixels, as we will see later,
the intensities measured still correspond to sampled values of a
new bandwidth limited function at the pixel grid. Treating the
new bandwidth limited function as an effective PSF, we can reconstruct it
by sampling theorem.
In fact, experiments found that the leading order intra-pixel variation
is common for all the pixels\cite{prfDef1}.
For micro-pixel level centrioding, however, we can not ignore
the inter-pixel variations of $Q_{mn}(x,y)$.
The pixel counts no longer strictly correspond to sampled values of
a bandwidth limited function. Fortunately, the inter-pixel variation is
only a small fraction of the total response, whose Fourier transform
can be characterized by laser metrology fringe measurements.

To illustrate this, we rewrite the pixel intensity model as
\beqa
  \!\!\!I_{mn} (x_c, y_c) &\!\!\!\! = & \!\!\!\!
 \int_{-\infty}^\infty \!\!\! dx \! \infint\!\!\! dy 
  \! \infint\! dk_x \! \infint \! dk_y {\cal I} (k_x, k_y)
 e^{i \left (k_x(x{-}x_c){+} k_y(y{-}y_c) \right )} Q_{mn}(x,y)
\nonumber\\
  &\!\!\!\! = & \!\!\!\! \infint dk_x \! \infint dk_y {\cal I} (k_x, k_y)
  e^{i \left (k_x((m{+}1/2)a{-}x_c){+}k_y((n{+}1/2)a{-}y_c) \right )}
\nonumber\\
 &\!\!\!\! & \quad \infint\!\! dx \infint dy
  Q_{mn} \!\! \left ((m{+}1/2)a{+}x,(n{+}1/2)a{+}y \right )
  \, e^{ i(k_xx {+}k_y y)}
\nonumber\\
  &\!\!\!\! = & \!\!\!\! \infint \! dk_x \!\! \infint\!dk_y
  {\cal I} (k_x, k_y) {\tilde Q}_{mn}\!(k_x, k_y)
  e^{i\left [k_x((m{+}1/2)a{-}x_c){+}k_y((n{+}1/2)a{-}y_c) \right ]}
\label{pixModelQeFt}
\eeqa
where $ {\cal I} (k_x, k_y)$ is the Fourier transform of $I(x,y)$ and
${\tilde Q}_{\rm mn}(k_x, k_y)$ is defined by
\beq
 {\tilde Q}_{\rm mn}(k_x, k_y) \equiv \infint dx \infint dy
   \; Q_{\rm mn} \left ((m{+}1/2)a{+}x,(n{+}1/2)a{+}y \right )
   \, e^{ i(k_xx {+}k_y y)}  \,.
\label{Q_mn_FT_def}
\eeq
The pixel response function related effects are all captured by
the Fourier transform ${\tilde Q}_{\rm mn}(k_x, k_y)$.

For the case that all pixels have identical response function,
\beq
  Q_{\rm mn} \left ((m{+}1/2)a{+}x,(n{+}1/2)a{+}y \right )
   = Q_{\rm 00} \left (a/2{+}x,a/2{+}y \right ) \,;
\eeq
and $Q_{\rm mn}(k_x, k_y)$ no longer depends on indices $m,n$,
\beq
  Q_{\rm mn}(k_x, k_y) = Q_{00}(k_x,k_y) \,.
\eeq
Putting this in the expression~(\ref{pixModelQeFt}), we obtain
\beq
  I_{mn}(x_c, y_c) = \infint dk_x \infint dk_y {\cal I} (k_x, k_y)
  Q_{00}(k_x, k_y)
  e^{i \left (k_x((m{+}1/2)a-x_c) + k_y((n{+}1/2)a-y_c) \right )} \,.
\eeq
Defining an effective PSF ${\bar I}(x,y)$
\beq
  {\bar I} (x, y) \equiv \infint dk_x \infint dk_y {\cal I} (k_x, k_y)
  Q_{00}(k_x, k_y) e^{i \left (k_x x + k_y y \right )}  \,,
\eeq
$I_{mn}(x_c, y_c)$ corresponds to values of function ${\bar I}(x,y)$
at grid points $(x,y){=}\left ((m{+}1/2)a{-}x_c,(n{+}1/2)a{-}y_c \right )$,
$m,n = 0, \pm 1, \pm 2, \cdots $.
Because ${\cal I}(k_x,k_y)$ vanishes for $|k_x| \ge 2\pi D/(\lambda f)$
or $|k_y| \ge 2\pi D/(\lambda f)$, ${\bar I}(x,y)$ is bandwidth limited
and the sampling theorem is applicable for reconstructing the
effective PSF ${\bar I}(x,y)$ using $I_{mn}(x_c, y_c)$.

In general, response function $Q_{mn}(x,y)$ depends on $(m,n)$. It is useful
to define an average response function ${\bar Q}(x,y)$
\beq
  {\bar Q} (x,y) \equiv {1 \over N^2}
    \sum_{m,n} Q_{mn}((m+1/2)a + x, (n+1/2)a+y) \,.
\eeq
and its Fourier transform
\beq
  {\tilde {\bar Q}} (k_x, k_y) \equiv \int dx \int dy {\bar Q}(x,y)
  e^{i \left (k_x x + k_y y \right )} \,.
\eeq
Because the common intra-pixel response function does not affect
the property of being sampled values of a bandwidth limited signal,
it is convenient to factor out the common pixel response to parametrize
the pixel dependent terms as
\beqa
  \!\!\!\!\!\!\!\! {\tilde Q}_{mn}(k_x,k_y) / {\tilde {\bar Q}} (k_x, k_y)
 &\!\!\! = &\!\!\!
  q_{mn} e^{ik_x\Delta x_{mn} + ik_y \Delta y_{mn}} 
  \left [1 + \alpha_{mn} k_x^2 + \beta_{mn} k_y^2 + \gamma_{mn} k_x k_y \right .
\nonumber \\
  && \left . \qquad \qquad \qquad \qquad \quad
  \;\;\;\, + \,a_{mn}k_x^3 + b_{mn} k_x^2 k_y + c_{mn} k_x k_y^2 + d_{mn} k_y^3
  + \cdots \right ] \,,
\label{QE_FT_exp}
\eeqa
where $q_{mn}$ represents the flat field response of pixel $(m,n)$,
$\Delta x_{mn}$ and $\Delta y_{mn}$ represent effective geometric
pixel location shifts along x and y directions for pixel $(m,n)$
to deviate from a regular pixel grid location $((m+1/2)a, (n+1/2)a)$.
$\alpha_{mn},\beta_{mn},\gamma_{mn}$ are parameters specifying
quadratic behavior in the amplitude of $Q_{mn}(k_x,k_y)$ that are pixel
dependent. $a_{mn},b_{mn},c_{mn},d_{mn}$ are third order term coefficients.
This parametrization is based on a Tayler series expansion
of ${\tilde Q}_{mn}(k_x,k_y)$ in terms of polynomials of $k_x$ and $k_y$.

\subsection{Centroid displacement estimation algorithm}
We now turn to present the algorithm for
estimating the centroid displacement between two images.
The main idea is to resample one of the two images 
at a grid shifted from the default grid by an offset.
When the resampled image best matches the second image,
the offset at which we resampled the first image is the estimated
centroid displacement. We call the first image (the image to be resampled)
 {\it reference} image.

Based on pixel grids, we use the following discrete frequencies
\beq
  (k_j, k_l) = {2 \pi \over Na} (j, l) \,,
  j = -N/2, \cdots, 0, 1, N/2-1 \,,\; l = -N/2, \cdots, 0, 1, N/2-1 \,.
\label{freqs}
\eeq
With this, our model is expressed as
\beq
  I_{mn}(x_c, y_c) = 
  \sum_{j=-N/2}^{N/2-1} \sum_{l=-N/2}^{N/2-1}
  {\cal M}_{jl} {{\tilde Q}_{mn}(k_j,k_l) \over {\tilde {\bar Q}}(k_j, k_l)}
  e^{i k_j \left [(m{+}1/2)a{-}x_c \right ]+
     i k_l \left [(n{+}1/2)a{-}y_c \right ]}
\label{discreteFreqModel}
\eeq
where
\beq
  {\cal M}_{jl} = {\cal I}(k_j, k_l) {\tilde {\bar Q}}(k_j, k_l) \,.
\label{effPsfFT}
\eeq 
Pixel response calibration measures
$ {\tilde Q}_{mn}(k_x,k_y) /{\tilde {\bar Q}}(k_x, k_y)$
at a set of spatial frequencies covering the bandwidth of PSF.
It is convenient to parametrize
$ {\tilde Q}_{mn}(k_x,k_y) /{\tilde {\bar Q}}(k_x, k_y)$
using low order terms in expression~(\ref{QE_FT_exp}).
If keeping the terms up to the third power,
we have then parameters $q_{mn}$, $\Delta x_{mn}, \Delta y_{mn}$,
$\alpha_{mn}, \beta_{mn}, \gamma_{mn}$,
$a_{mn}, b_{mn}, c_{mn}, d_{mn}$. They
can be estimated by fitting expansion~(\ref{QE_FT_exp})
to calibration measurements, which are values at a set of spatial
frequencies pre-determined by the metrology system for pixel
response calibration.

To reconstruct the PSF, we need to estimate ${\cal M}_{jl}$.
Let us choose the coordinate so that the centroid of the reference image is
at the origin, {\it i.e.}~$x_c = 0,y_c=0$ and its pixel intensities
are represented by $I^0_{mn}$. Detector calibration measures flat pixel
responses $q_{mn}$, effective pixel offsets $\Delta x_{mn}$,
$\Delta y_{mn}$, quadratic
amplitude coefficients $\alpha_{mn}, \beta_{mn}$ and $\gamma_{mn}$,
and the third order phase parameters $a_{mn}, b_{mn}, c_{mn}, d_{mn}$.
Relation~(\ref{discreteFreqModel})
\beq
  I_{mn}^0 = 
  \sum_{j=-N/2}^{N/2-1} \sum_{l=-N/2}^{N/2-1}
  {\cal M}_{jl} {{\tilde Q}_{mn}(k_j,k_l) \over {\tilde {\bar Q}}(k_j, k_l)}
  e^{i k_j (m{+}1/2)a+ i k_l (n{+}1/2)a}
\eeq
can be inverted to solve for ${\cal M}_{jl}$, which is the frequency
representation of the image.
We use expression~(\ref{discreteFreqModel}) to resample the reference
image. The centroid displacement
of the second image $I_{mn}$ relative to the first image can be estimated by
solving a weighted nonlinear least-squares fitting as
\beq
  (x_c, y_c) = \min_{x_c, y_c} \sum_{mn} \left |
  I_{mn} - I_{mn}(x_c, y_c) \right |^2 W_{mn} 
\eeq 
where $W_{mn}$ is weight factor for pixel $(m,n)$.
The choice of $W_{mn}$ affects sensitivity to noise.
Here we simply use equal weights and save
optimizing the weights to achieve best noise sensitivity
as a future topic.

When we use discrete frequencies~(\ref{freqs}), the signal
is parametrized as periodical function in space with period $Na$.
For $N=32$, the image encircles the 7th Airy ring; the PSF is
small enough at the boundary so that the truncation error
is negligible.

\section{Pixel response calibration using laser metrology}
\label{sec:metCal}
We measure the pixel response functions in Fourier space
by observing the response of the detector to a sinusoidal intensity
illumination pattern. The sinusoidal pattern is generated by fringes
from interfering two laser metrology beams far away so
that the wavefront is close to a plane wave over the spatial
extension of the detector. To make the identification of the laser
fringes easier, we use Acoustic Modulation Oscillator (AMO) to
offset one of the laser's frequency from
the other by a few Hz so that the fringes move across the CCD.
Therefore, the illuminating intensity may be modeled as
\beq
  I(x, y, t) =  I_1^{\rm met} + I_2^{\rm met} +
  2 \sqrt{I_1^{\rm met} I_2^{\rm met}} {\rm Re} e^{i(k_x^{\rm met} x
  + k_y^{\rm met} y + \Delta \omega t)} \,,
\eeq
where $I_1^{\rm met}$ and $I_2^{\rm met}$ are the intensities
of the two lasers, $(k_x^{\rm met}, k_y^{\rm met})$ is
the spatial wave number of the laser fringes, and $\Delta \omega$
is the angular frequency difference between the two lasers introduced
by the AMO. The output counts of pixel $(m,n)$ is then
\beqa
  \! \!\!\!\!\! I_{\rm mn}^{\rm met}(t) &\!\!\!\! =  &\!\!\!\!
  \int dx \int dy \, I(x,y,t) Q_{\rm mn}(x,y)
\nonumber\\
  &\!\!\!\! = &\!\!\!\!
  (I_1^{\rm met}{+}I_2^{\rm met}){\tilde Q}_{\rm mn}(0,0){+}
  2 \sqrt{I_1^{\rm met} I_2^{\rm met}} {\rm Re} \! \left \{ \!
  {\tilde Q}_{\rm mn}(k_x^{\rm met} \!, k_y^{\rm met})
  e^{i(k_x^{\rm met} (m{+}1/2)a{+}k_y^{\rm met}(n{+}1/2)a{+}\Delta \omega t)}
  \! \right \} ,
\eeqa
where we have used the definition~(\ref{Q_mn_FT_def}).
The temporal variation of the intensity at each pixel is a sinusoidal
function plus a constant. By estimating the amplitude and phase of the
sinusoidal temporal variation, we get the complex Fourier transform
${\tilde Q}_{\rm mn}(k_x^{\rm met}, k_y^{\rm met})$ because
the overall intensities of the two metrology beams $I_1^{\rm met}$
and $I_2^{\rm met}$ can be easily measured.
By having different separations between the laser beams and distances
between the laser source and the CCD, we can generate fringes
of different spatial frequencies and thus measure
${\tilde Q}_{\rm mn}(k_x^{\rm met}, k_y^{\rm met})$ at various values
of $(k_x^{\rm met}, k_y^{\rm met})$.
Note that we only need to measure the Fourier transform of the
the pixel response functions to the highest frequency in
the PSF, which is limited by the aperture size of the optics.
For Nyquist sampled images, it is only necessary to measure
the spatial frequency to $\pi/a$, where $a$ is the spacing
between pixels.

\section{Results using simulated data}
We use simulated data to validate our concept of the algorithm.
In subsection~\ref{sec:idealCCD},
we first show that for an ideal detector whose pixels have
the same response function, we can achieve sub-micro-pixel level accuracy in
centroid estimation. 
\subsection{Results for an ideal detector}
\label{sec:idealCCD}
\begin{figure}
\begin{center}
\begin{tabular}{c}
\includegraphics[height=6.5cm]{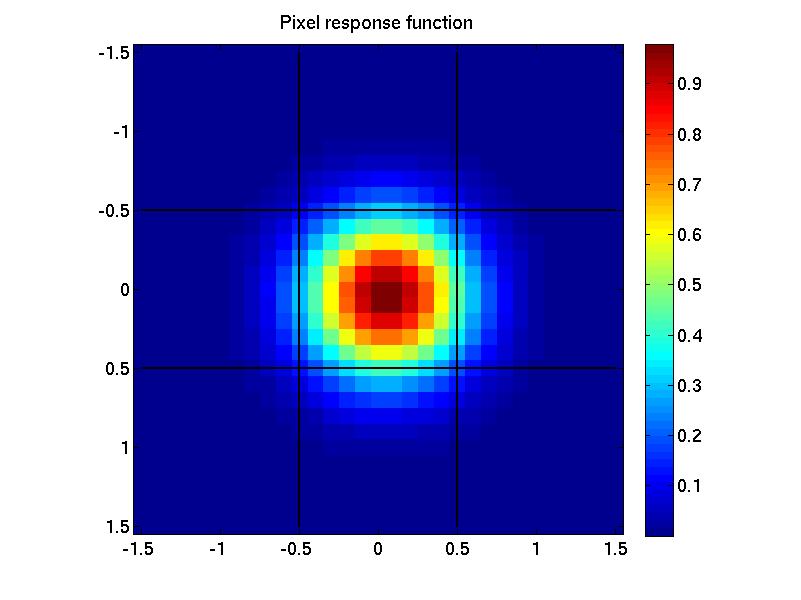}
\includegraphics[height=6.5cm]{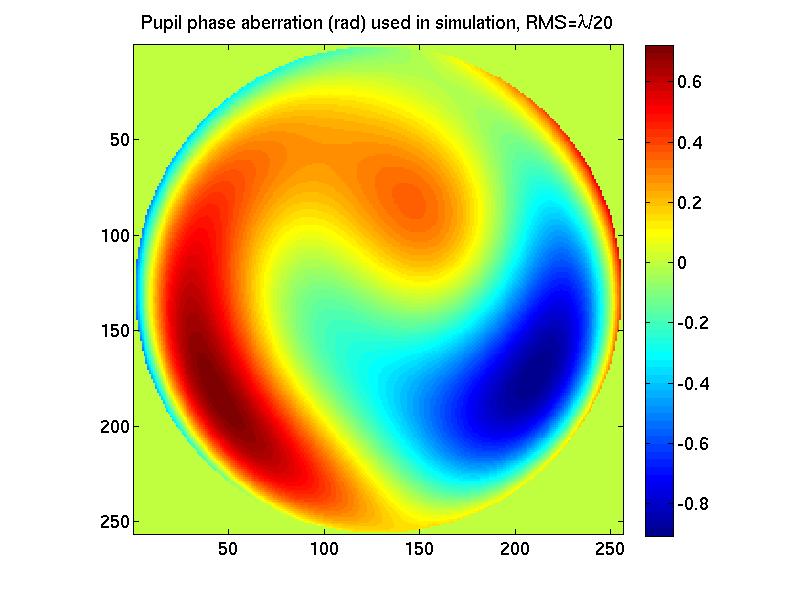}
\end{tabular}
\end{center}
\caption{A common pixel response function for all the pixels used in the simulation
(left) and low order phase aberration (right) in radian used in simulation with
RMS $\sim \lambda/20$.} 
\label{commonPixRespAndWfe}
\end{figure}

\begin{figure}
\begin{center}
\begin{tabular}{c}
\includegraphics[height=6.5cm]{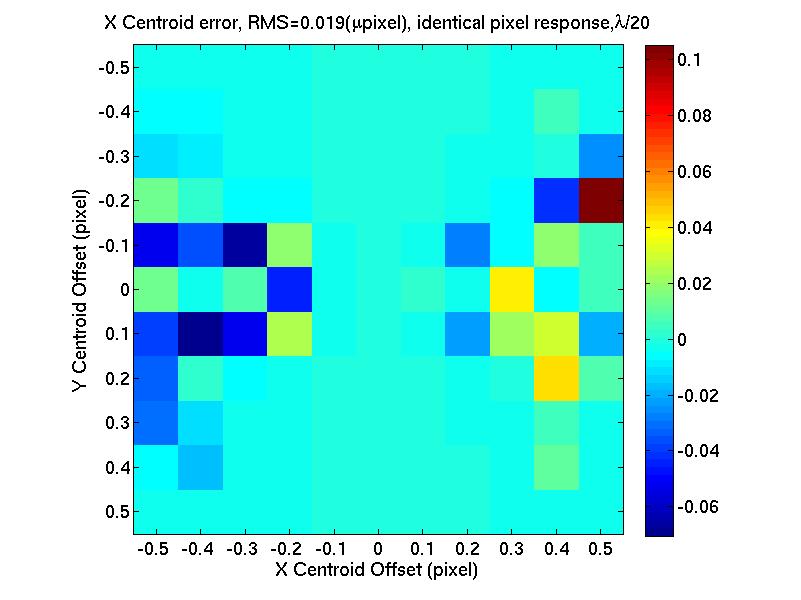}
\includegraphics[height=6.5cm]{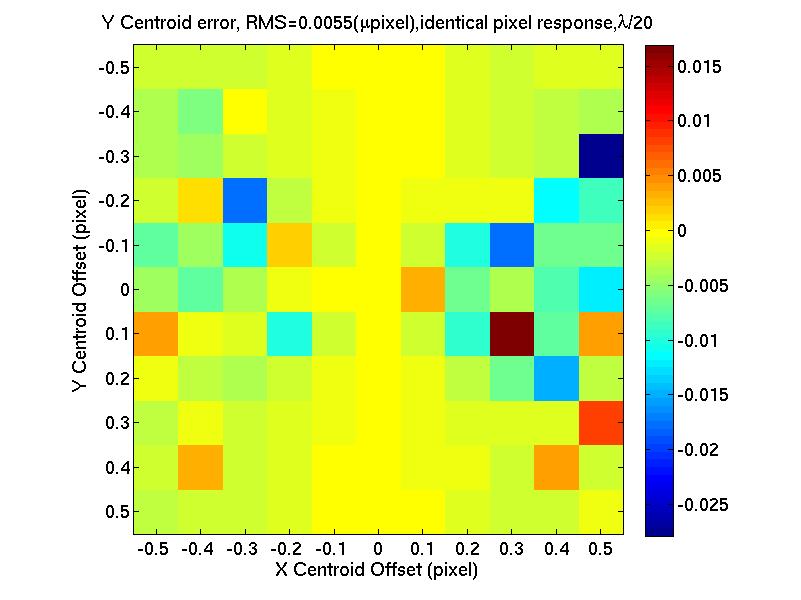}
\end{tabular}
\end{center}
\caption{Centroid offset estimation errors X (left) and Y (right)
in micro-pixel unit for an ideal CCD and wave front with $\lambda/20$ RMS
low order aberrations parametrized by the first 15 Zernike polynomials.
The pixel size is 10$\mu$m.}
\label{samePixResp_20th_WFE}
\end{figure}

In our simulation, the diameter of the telescope $D$=1m with
focal length $f=40m$ to be consistent with NEAT\cite{NEAT}.
We focus on monochromatic source with the wavelength $\lambda = 600nm$.
In section~\ref{sec:polychrom}, we discuss the case of polychromatic source.
We only consider an array of 32$\times$32 pixels, i.e. the dimension N=32
because this is sufficient for achieving micro-pixel centroid estimation.
The corresponding $\lambda/D$ at the focal plane is 24$\mu$m.
The pixel size is 10$\mu$m, which samples above the Nyquist frequency (1/(12$\mu$m)).
All the pixels have the same pixel response function, which is displayed
in the left plot in Fig.~\ref{commonPixRespAndWfe}.
It is modeled as a Gaussian function
multiplied by low order polynomials (up to 4th order),
\beq
  Q(x,y) = \exp(-(x^2+y^2)/r_g^2) \left [c_0 + c_1 x + c_2 y + 
  c_3 x^2 + c_4 y^2 + c_5 x y + \cdots \right ]
\label{pixelRespModel}
\eeq
where $x,y$ are coordinates in the detector plane in unit of pixel,
$ r_g = 0.5$, and the coefficients $c_i$ are drawn from
Gaussian random number generators with standard deviation being 0.05.
The mean values are all 0 except for $c_0$ whose mean is 0.8.
This pixel response function is roughly similar to the the intra-pixel variation
for a backside-illuminated CCD\cite{intraBackIllu}.
We include $\lambda/20$ RMS low order wavefront aberrations
parametrized by the first 15 Zernike polynomials, whose amplitudes are
randomly generated. The right plot in Fig.~\ref{commonPixRespAndWfe}
displays the phase aberration over the telescope pupil used in our simulation.
The two plots (left for X centroid and right for Y centroid) in
Fig.~\ref{samePixResp_20th_WFE} displays the centroid estimation errors
for a grid of X and Y offsets between the two
images within range [-0.5, 0.5] pixel. These errors are
due to truncation and are no more than 0.1 micro-pixel.

\subsection{Pixel response calibration and results}
For realistic detectors, pixel response functions vary
from pixel to pixel, which we call {\it inter-pixel} variations.
Pixel response calibration measures the pixel response
functions of all the pixels. The Fourier transforms
${\tilde Q}_{mn}(k_x, k_y)$ of the pixel response functions
$Q_{mn}(x,y)$ of all pixels are measured at various spatial frequencies.
To do this, we illuminate the detector using a cosine intensity pattern.
For NEAT, this is achieved by interfering two metrology
lasers at various spatial frequencies . The fringe pattern gives
a sinusoidal illumination on the pixel array
and the separation of the two lasers
determines the wavelengths of the fringes. AMO is used to offset the
frequency of one laser from the other by a few Hz. The temporal variation
of the pixel intensities are then used to
estimate Fourier transforms ${\tilde Q}_{mn}(k_x, k_y)$
as discussed in section~\ref{sec:metCal}.
We use expansion~(\ref{QE_FT_exp}) to parametrize ${\tilde Q}_{mn}(k_x, k_y)$.
For micro-arcsecond accuracy, we nominally keep the terms up to second order
or third order in $k_x$ and $k_y$.

\begin{figure}
\begin{center}
\begin{tabular}{c}
\includegraphics[height=6.5cm]{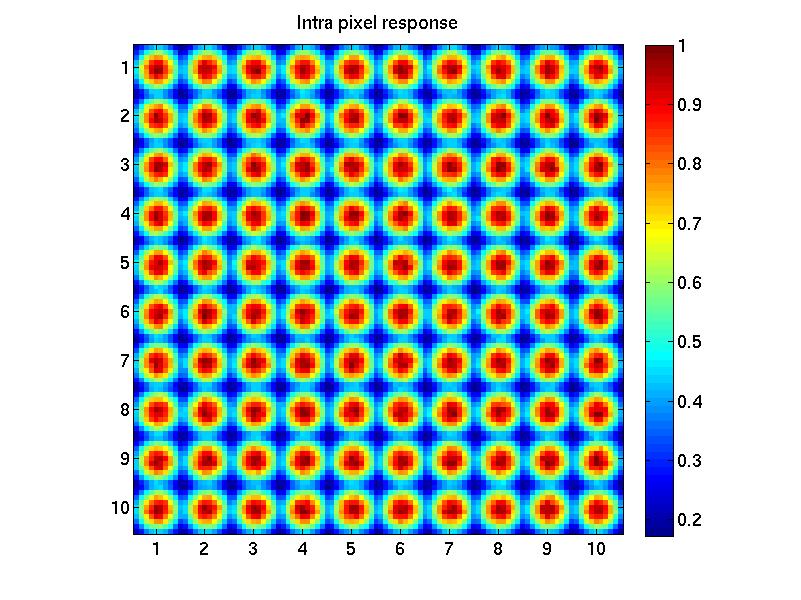}
\includegraphics[height=6.5cm]{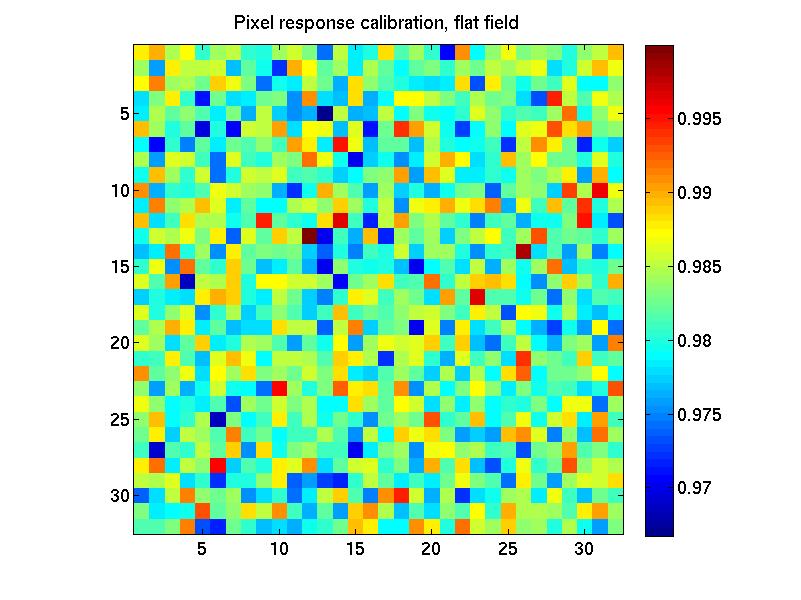}
\end{tabular}
\end{center}
\caption{Simulated intrapixel detection variation for a 10$\times$10 array (left)
and flat field response calibration result 32$\times$32 (right).}
\label{intraPixVar10x10_FlatQE}
\end{figure}

\begin{figure}
\begin{center}
\begin{tabular}{c}
\includegraphics[height=6.5cm]{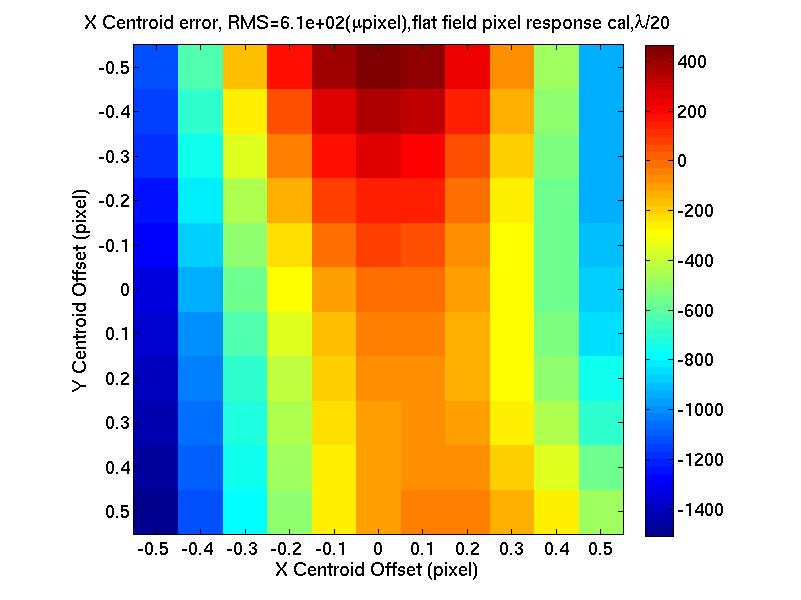}
\includegraphics[height=6.5cm]{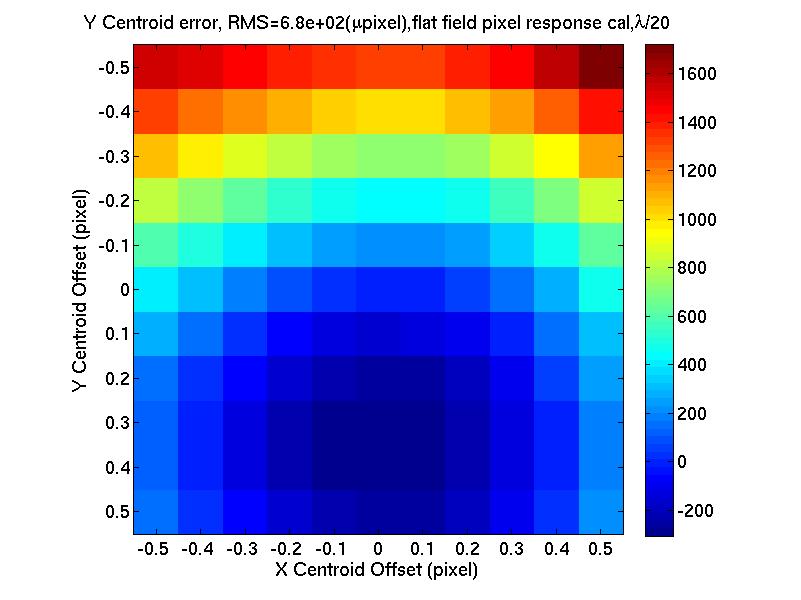}
\end{tabular}
\end{center}
\caption{Centroid offset estimation errors X (left) and Y (right) in micro-pixel unit. The wavefront error is $\lambda/20$ specified by the first 15 Zernikes polynomials. Only a flat field calibration is applied.}
\label{centErrsFlatFieldCal}
\end{figure}

\begin{figure}
\begin{center}
\begin{tabular}{c}
\includegraphics[height=6.5cm]{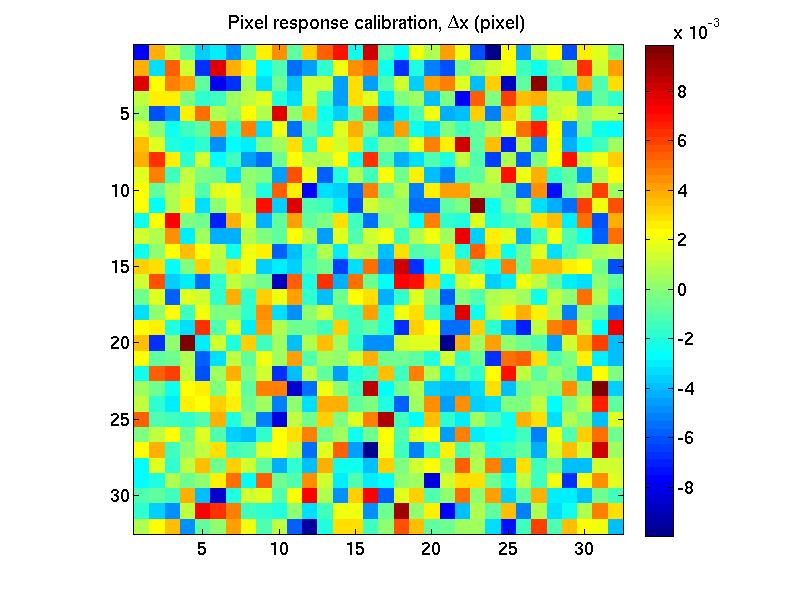}
\includegraphics[height=6.5cm]{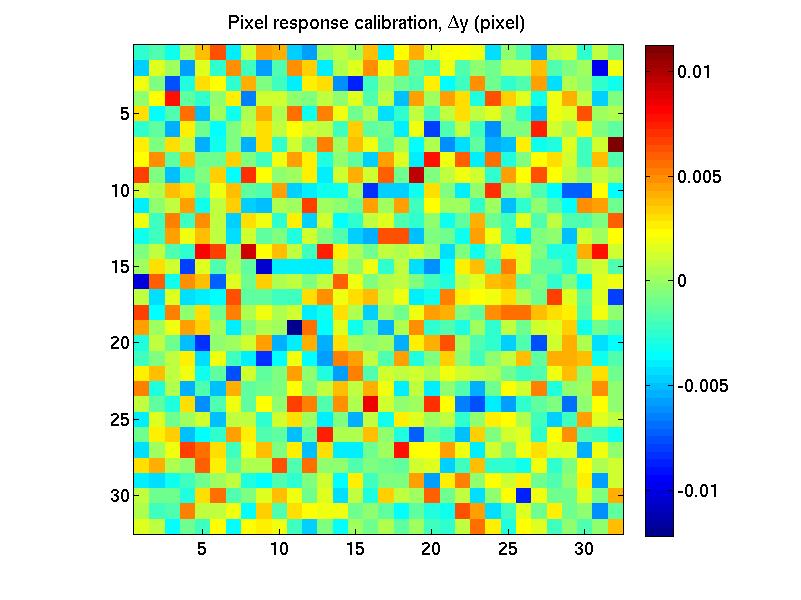}
\end{tabular}
\end{center}
\caption{Effective pixel offsets along x (left) and y directions, the units are pixel}
\label{pixelOffsets}
\end{figure}

\begin{figure}
\begin{center}
\begin{tabular}{c}
\includegraphics[height=6.5cm]{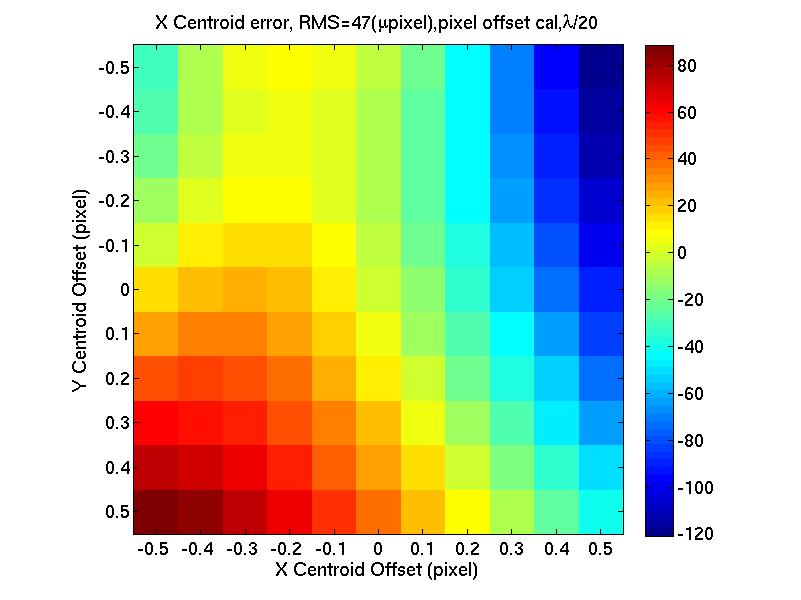}
\includegraphics[height=6.5cm]{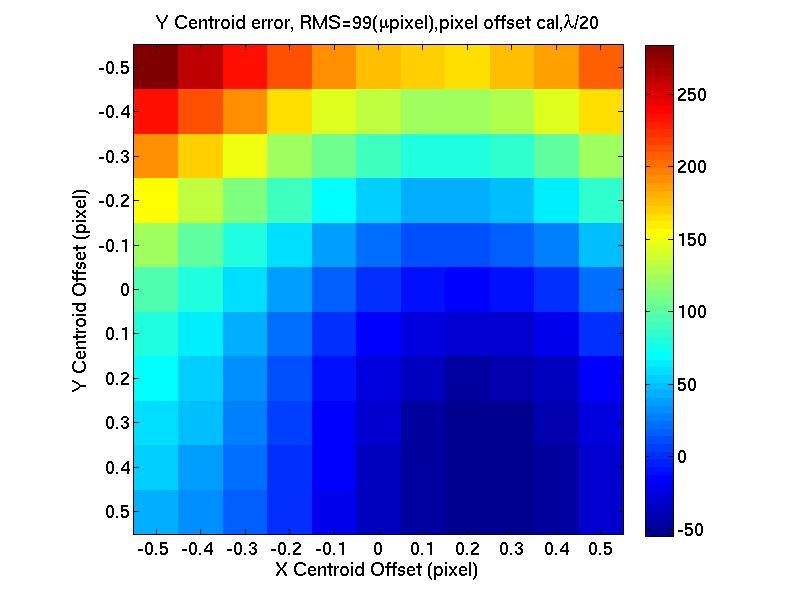}
\end{tabular}
\end{center}
\caption{Centroid offset estimation errors X (left) and Y (right) in micro-pixel unit. The wavefront error is $\lambda/20$ specified by the first 15 Zernikes polynomials.
 A flat field and pixel effective location calibration is applied.}
\label{pixelOffsetCal}
\end{figure}

\begin{figure}
\begin{center}
\begin{tabular}{c}
\includegraphics[height=4.5cm]{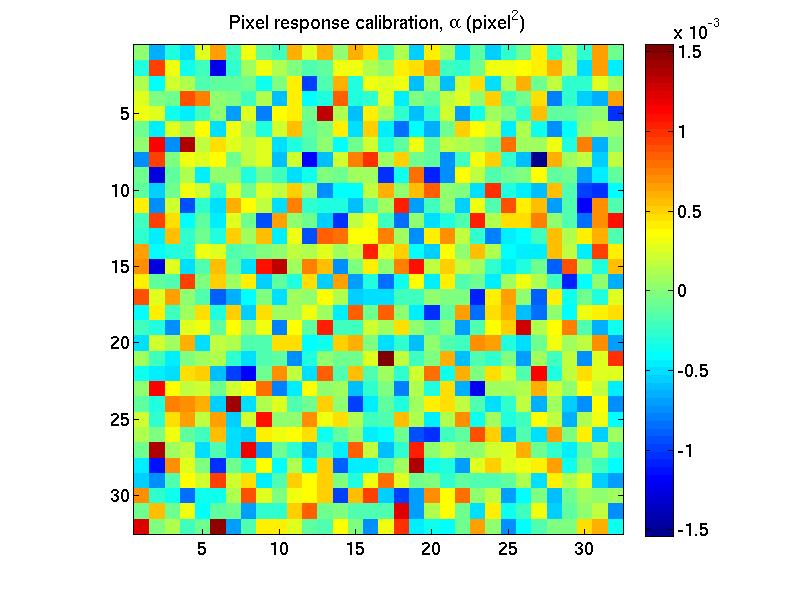}
\includegraphics[height=4.5cm]{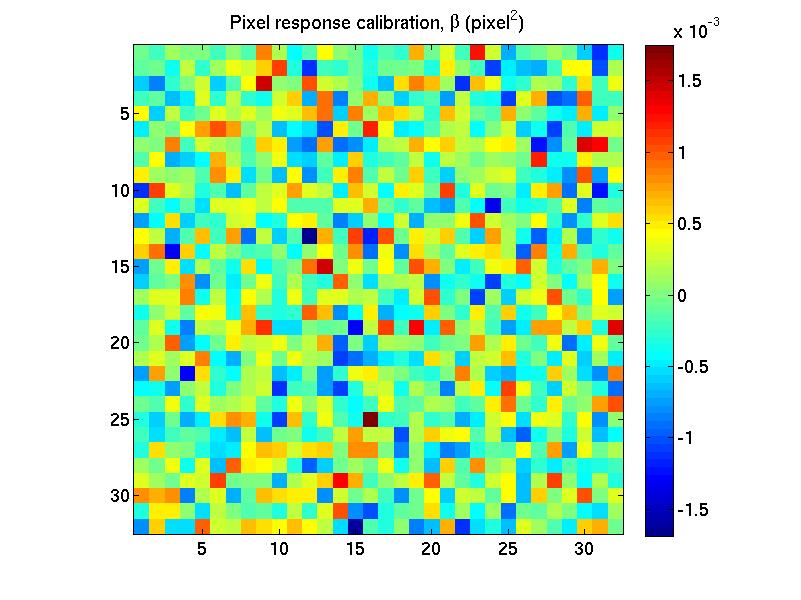}
\includegraphics[height=4.5cm]{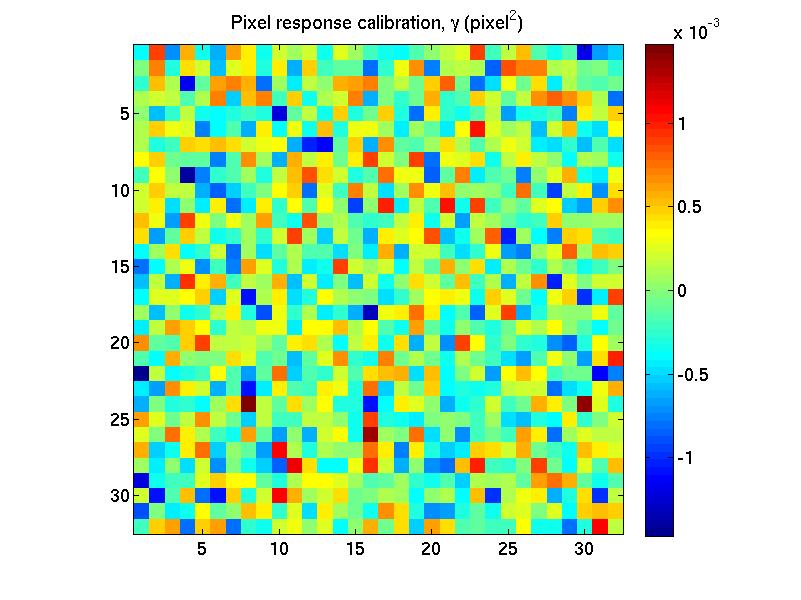}
\end{tabular}
\end{center}
\caption{Second order amplitudes $\alpha_{mn}$ (left), $\beta_{mn}$ (middle),
and $\gamma_{mn}$ (right), coefficients for terms $k_x^2, k_y^2$, and $k_x k_y$.}
\label{2ndAmp}
\end{figure}

\begin{figure}
\begin{center}
\begin{tabular}{c}
\includegraphics[height=6.5cm]{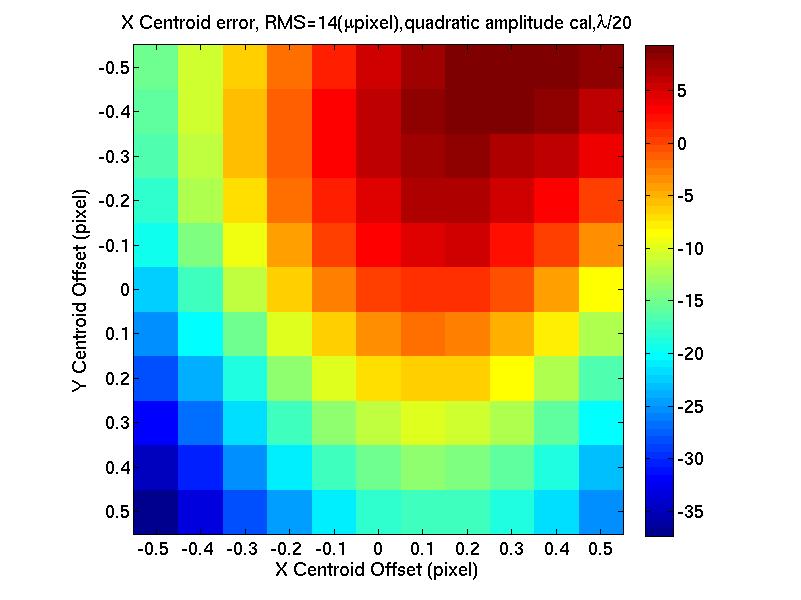}
\includegraphics[height=6.5cm]{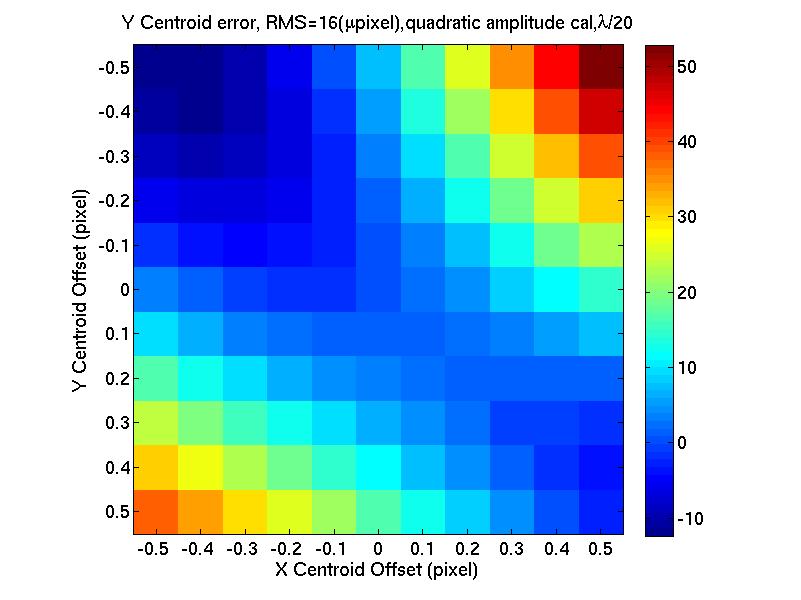}
\end{tabular}
\end{center}
\caption{Centroid offset estimation errors X (left) and Y (right) in micro-pixel unit. The wavefront error is $\lambda/20$ specified by the first 15 Zernikes polynomials. Calibration includes second order terms.}
\label{2ndAmpCal}
\end{figure}

\begin{figure}
\begin{center}
\begin{tabular}{c}
\includegraphics[height=5.5cm]{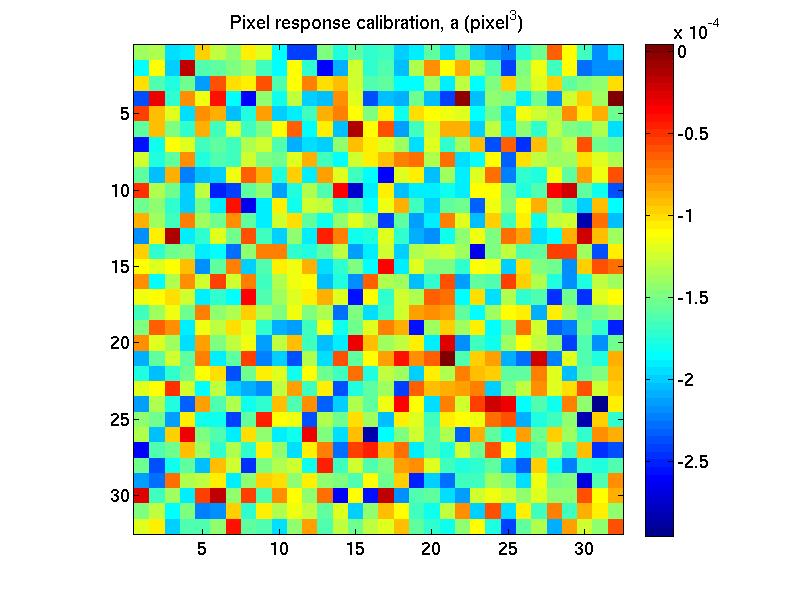}
\includegraphics[height=5.5cm]{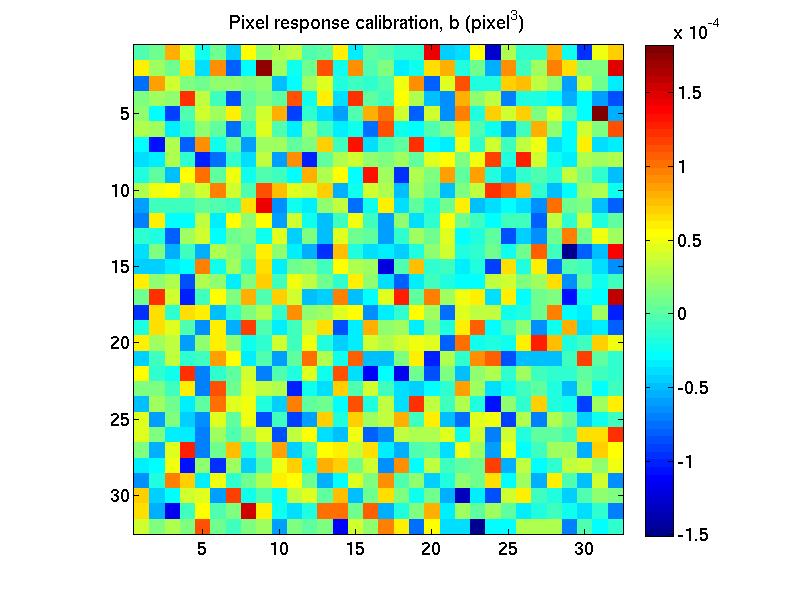}
\\
\includegraphics[height=5.5cm]{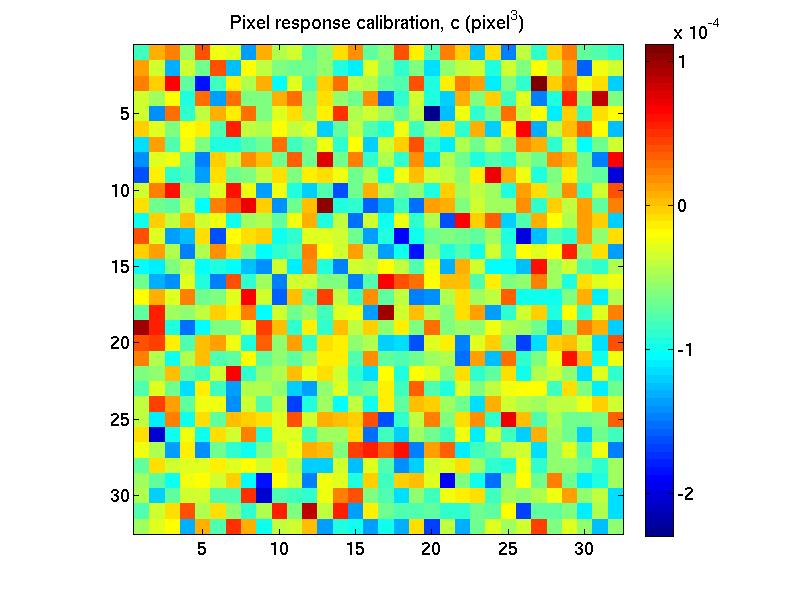}
\includegraphics[height=5.5cm]{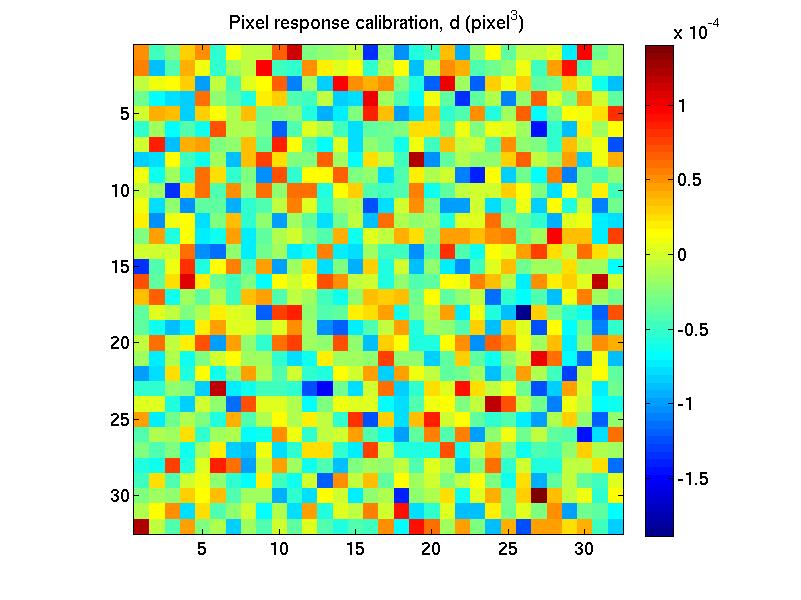}
\end{tabular}
\end{center}
\caption{Third order term coefficients, $a k_x^3$ (top, left), $b k_x^2 k_y$ (top, right), $c k_x k_y^2$ (bottom, left), and $d k_y^3$ (bottom, right).}
\label{3rdOrderPhase}
\end{figure}

\begin{figure}
\begin{center}
\begin{tabular}{c}
\includegraphics[height=6.5cm]{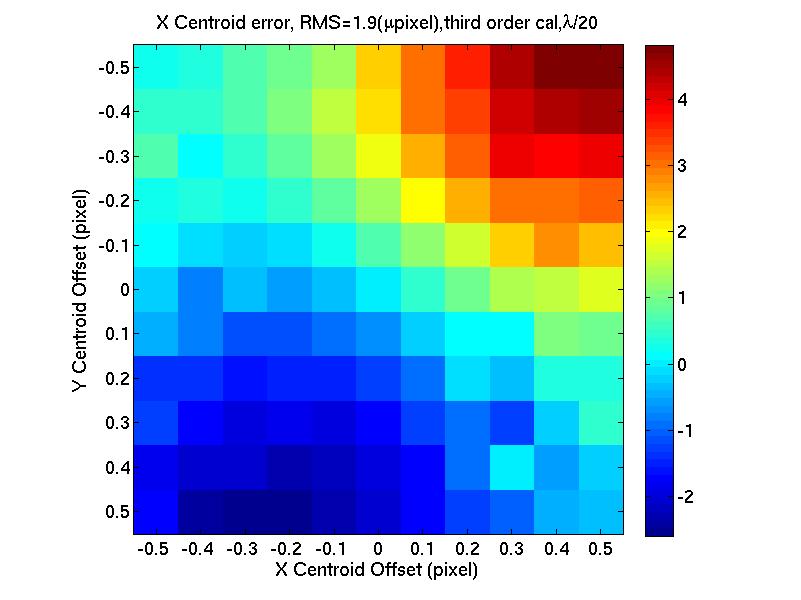}
\includegraphics[height=6.5cm]{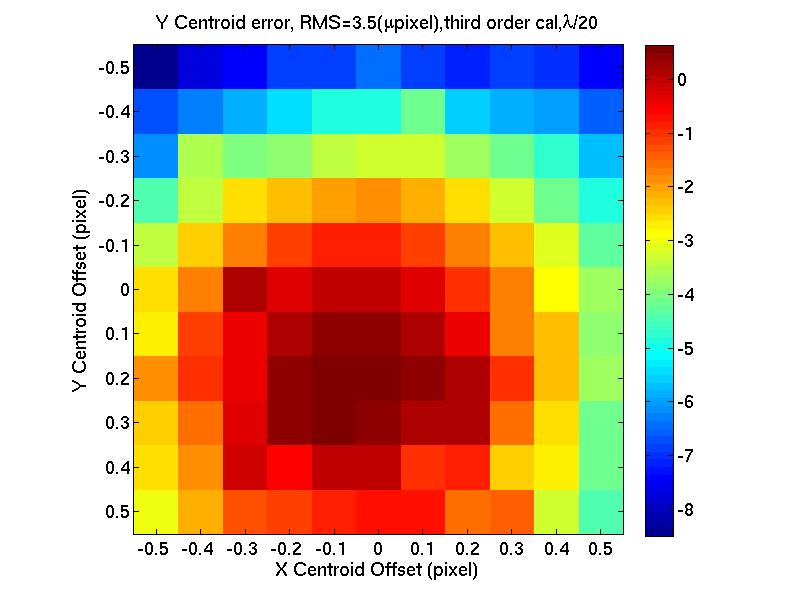}
\end{tabular}
\end{center}
\caption{Centroid offset estimation errors X (left) and Y (right) in micro-pixel unit. The wavefront error is $\lambda/20$ specified by the first 15 Zernikes polynomials.}
\label{perf_3rdOrderCal}
\end{figure}

To simulate the inter-pixel response variations,
we make the low order coefficients $c_i$ in Eq.~(\ref{pixelRespModel})
pixel dependent by replacing $c_i$ with $c_i + c_i^{\rm mn}$, where
$c_i^{\rm mn}$ are random numbers drawn from zero mean Gaussian random
number generators with standard deviation being 0.01. We also add
2\% white noise to the pixel response as a multiplicative factor
as well as random geometric pixel location shifts for all the pixels
along both x and y directions with 0.01 pixel RMS.
The left plot in Fig.~\ref{intraPixVar10x10_FlatQE} displays a typical
intra pixel responses of a 10$\times$10 pixel array at the upper left
corner of the 32$\times$32 array. Here the interpixel cross
talk from diffusion is not displayed to avoid overlap.
The zeroth order calibration measures the pixel response to a flat field,
{\it i.e.}~Fourier transform of the pixel response function at
spatial wave number $(k_x, k_y) = (0,0)$.
The right plot in Fig.~\ref{intraPixVar10x10_FlatQE} displays a typical flat
field calibration results.
Fig.~\ref{centErrsFlatFieldCal} shows systematic centroid displacement
estimation errors with only flat field response calibration for
different centroid offsets between the two images,
whose range is [-0.5, 0.5] pixel along both x and y directions. 
The next level of calibration measures the effective pixel locations deviating
from a regular grid.
The effective location of pixel $(m,n)$ is estimated by fitting
$\Delta x_{mn} k_x^{\rm met} + \Delta y_{mn} k_y^{\rm met}$  to the phase
of the estimated Fourier transform
${\tilde Q}_{\rm mn}(k_x^{\rm met}, k_y^{\rm met})$ at different
values of spatial frequency $(k_x^{\rm met}, k_y^{\rm met})$ available from
the metrology system.
Fig.~\ref{pixelOffsets} displays the estimated the effective pixel location deviating
from a regular grid.
Now the errors shown in Fig.~\ref{pixelOffsetCal} are significantly reduced to tens of micro-pixels after including the effective pixel locations in estimation. 
However, the errors are still large for micro-pixel level astrometry.
We further include the second order amplitude terms in expansion~(\ref{QE_FT_exp}) and
estimate coefficients $\alpha_{mn}, \beta_{mn}$, and $\gamma_{mn}$ by fitting
expansion~(\ref{QE_FT_exp}) to the estimated Fourier transforms
${\tilde Q}_{\rm mn}(k_x^{\rm met}, k_y^{\rm met})$ at different
values of $(k_x^{\rm met}, k_y^{\rm met})$.
The second order coefficients are displayed in Fig.~\ref{2ndAmp}.
The two plots in Fig.~\ref{2ndAmpCal} displays
the centroid estimation errors for performing pixel response
calibration that estimates the flat field response, the effective pixel 
locations, and the second order amplitude corrections.
Including the third order terms in expansion~(\ref{QE_FT_exp}) in our
pixel response calibration enables us to achieve centroid estimation accuracy
to be a few micro-pixels over [0.5, 0.5] pixel range along both x and y
directions. The corresponding coefficients are displayed in Fig.~\ref{3rdOrderPhase}.
Fig.~\ref{perf_3rdOrderCal} shows systematic centroid displacement estimation errors
for including through the third order phase terms. The RMS is only a few micro-pixels.
\newpage

\subsection{Noise sensitivity}
\begin{figure}
\begin{center}
\begin{tabular}{c}
\includegraphics[height=6.5cm]{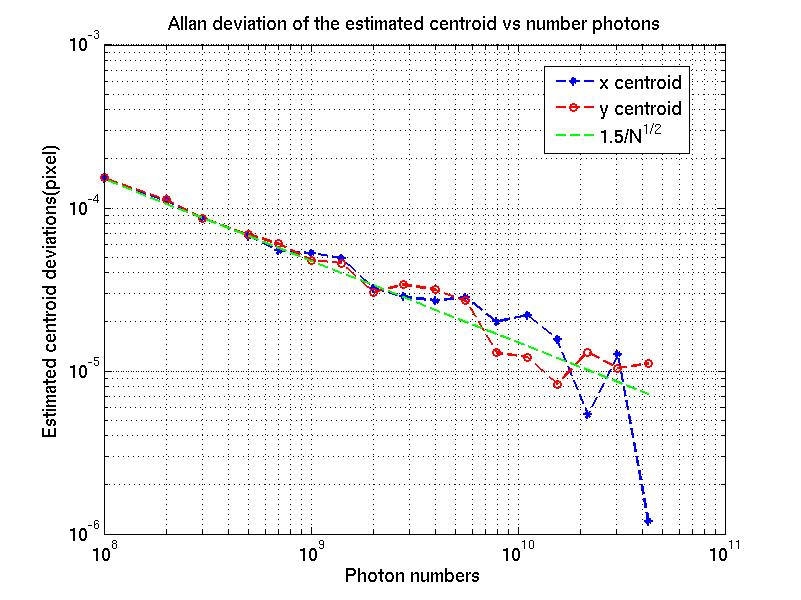}
\end{tabular}
\end{center}
\caption{Centroid estimation error Allan deviations vs number of photons integrated.}
\label{allanDev}
\end{figure}

\begin{figure}
\begin{center}
\begin{tabular}{c}
\includegraphics[height=6.5cm]{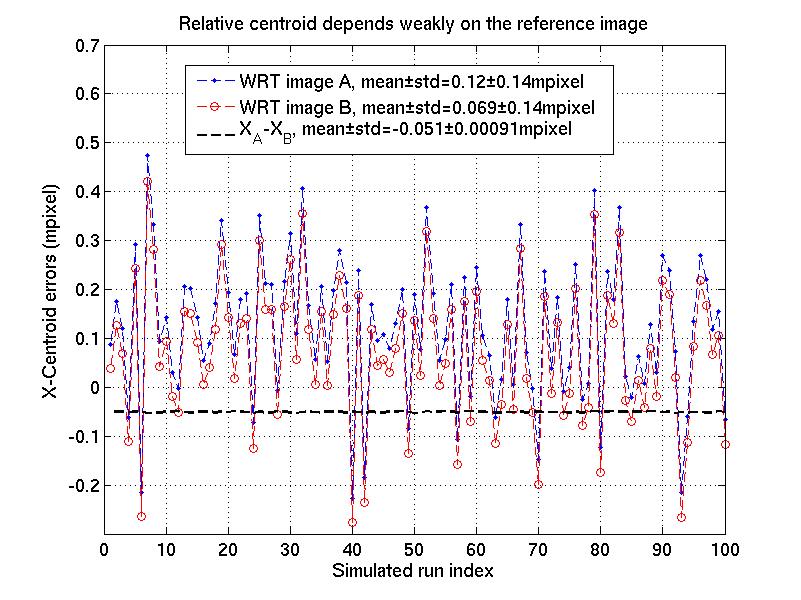}
\end{tabular}
\end{center}
\caption{X centroid displacement error using reference image with different noise
realizations.}
\label{weakRefDep}
\end{figure}

So far, we have not included any noise. We now study the sensitivity to
photon shot noise, which is the dominant source of random errors.
We set the total number of photons for each image to be $10^8$ and
use Poisson random number generator to simulate the photon shot noise.
We simulated 1000 images with their centroid positions randomly distributed within
$\pm$ 0.5 pixel relative to a reference image along both x and y directions.
The reference image have the same level of photon and shot noise. 
Fig.~\ref{allanDev} displays the Allan deviation of the estimated centroid
displacements relative to a reference image as function of the total number of
photons integrated. The green dash line shows an empirical sensitivity formula
\beq
  \sigma \approx 1.5/\sqrt{N_{\rm ph}}
\eeq
for uncertainty of centroid estimation using an equal weight in the least-squares
fitting, where $N_{\rm ph}$ is the total number of photons. 
To reach 10 mirco-pixel precision, we shall need about 2.2$\times$10$^{10}$ photons.
The Allan deviation shows the variation between 1000 images with respect to
the same reference image. How does the noise in the reference image affect
the results? It turns out that the noise in the reference image
causes mostly an overall offset to all the centroid displacements,
or it is an excellent approximation that 
\beq
  d(A, R_1) - d(B, R_1) \approx d(A, R_2) - d(B, R_2) \,,
\eeq
where $d(A,R_1)$ represents the displacement of centroid of image
A relative to reference image $R_1$ and so on.
Fig.~\ref{weakRefDep} shows the centroid displacements relative to two
reference images, which are the same image with two different realization
of photon noise. It is easy to see that
the difference of the centroid displacements
is insensitive to the uncertainties in the reference image( less than 1 micro-pixel).
We are performing an equal weight least-squares fitting,
it is possible to optimize weights $W_{mn}$ to achieve the best
sensitivity to noise, which will be a subject for future study.

\subsection{Wave front stability}
\begin{figure}
\begin{center}
\begin{tabular}{c}
\includegraphics[height=6.5cm]{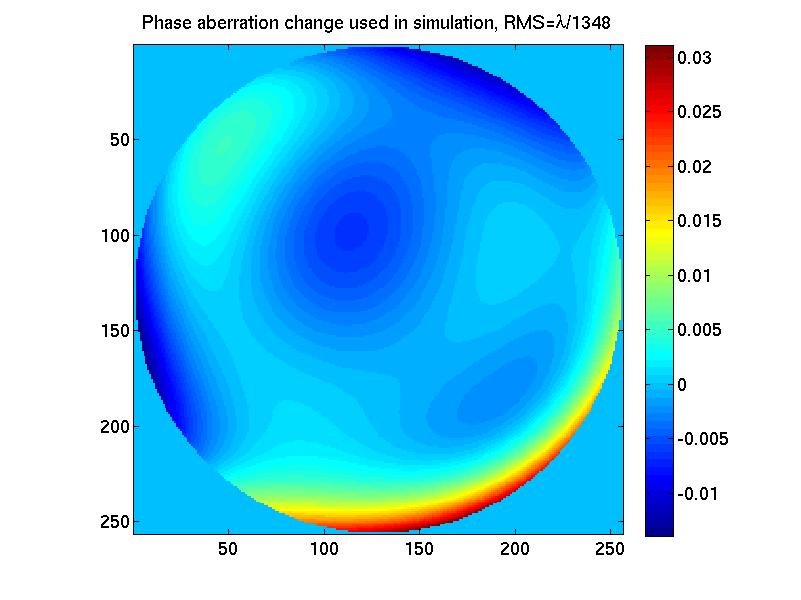}
\includegraphics[height=6.5cm]{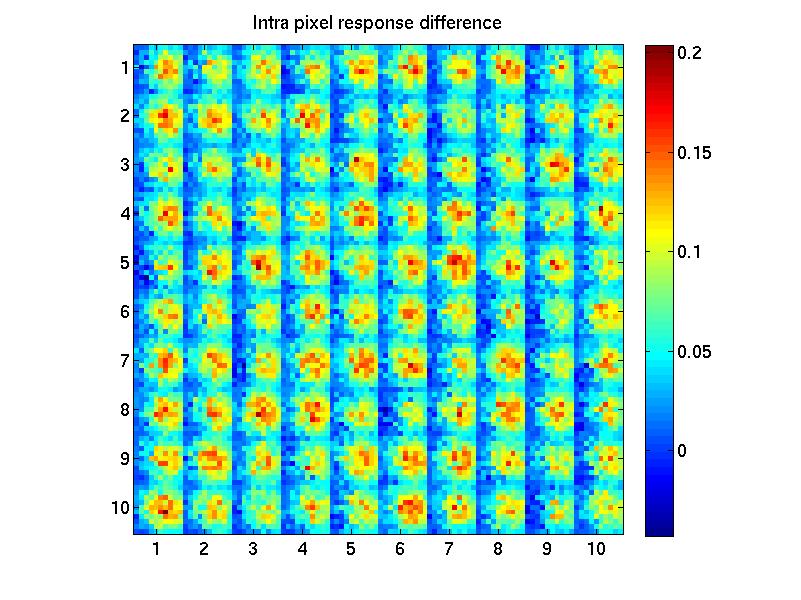}
\end{tabular}
\end{center}
\caption{Pixel response difference between two detectors in a 10$\times$10 array (left). Wavefront variations between two observations, RMS $\sim\lambda/1350$, simulated by low order polynomials with randomly generated coefficients.}
\label{PixRespDiffAndWfVar}
\end{figure}

In this section, we study the sensitivity of our algorithm to
wave front changes, i.e. the two images are taken with slightly
different wavefront aberrations. The left plot in Fig.~\ref{PixRespDiffAndWfVar}
shows a low order small wavefront change with $\lambda/1350$ RMS.
(High order wave front aberrations generates speckles in the image plane while
the low order wavefront change generates a change in the shape of the PSF.
For PSF fitting, the centriod estimation is more sensitive to low order
wavefront aberrations.) The main effect of a wavefront change
between the two images is an overall constant, which is not a problem
for differential astrometry. However, this overall constant depends
on the pixel response functions. We shall need to show that
this pixel dependency in the overall constant is small so that it is
common for different portions of the detector.
To do this, we instantiate a second detector whose pixel response function is
slightly different from our nominal detector as a result of the random
feature of our simulation.

\begin{figure}
\begin{center}
\begin{tabular}{c}
\includegraphics[height=6.5cm]{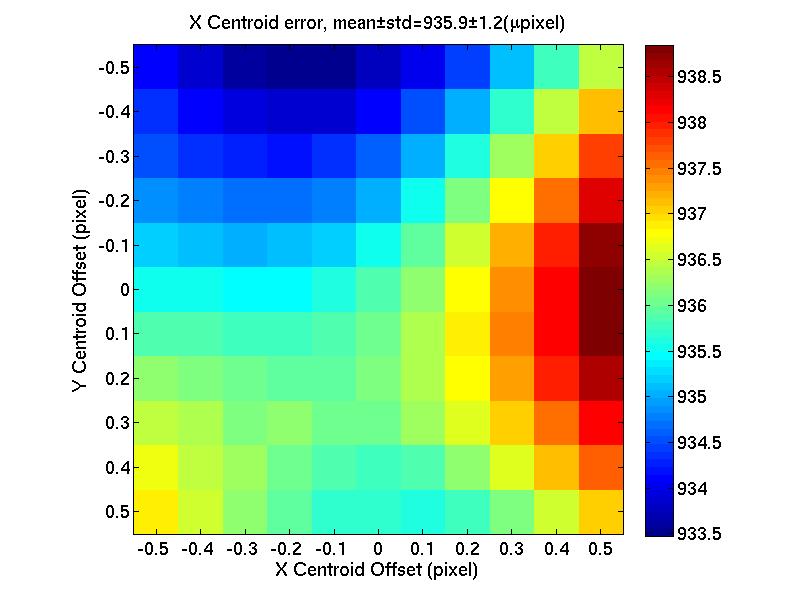}
\includegraphics[height=6.5cm]{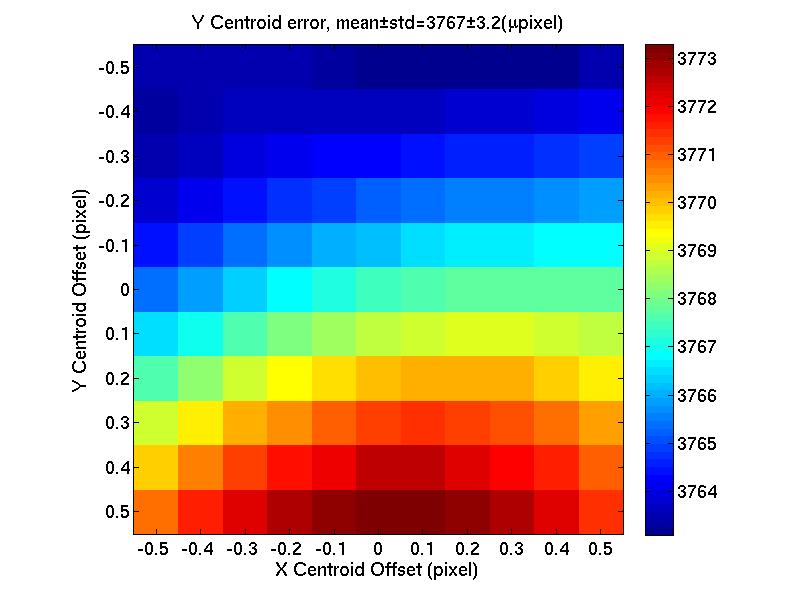}
\end{tabular}
\end{center}
\caption{Centroid displacement estimation errors X (left) and Y (right) in micro-pixel unit. Both images are taken with a common low order wavefront error of $\lambda/20$ RMS specified by the first 15 Zernikes polynomials. The low order wavefront difference between the two images has RMS $\sim\lambda/1350$.}
\label{wfChangeEffect}
\end{figure}

The right plot in Fig.~\ref{PixRespDiffAndWfVar} shows the difference of
the pixel responses between the two cameras which comes from
the randomly instantiated low order polynomials.
We now consider the case where the wavefront changes by
an amount of $\lambda/1350$ RMS, as shown in the left plot in
Fig.~\ref{PixRespDiffAndWfVar}, between taking the two images
whose centroid displacement we are interested to estimate.
Fig.~\ref{wfChangeEffect} displays the centroid displacement errors
due to the $\lambda/1350$ RMS low order wavefront change for a grid of
centroid offsets between the two images.
The main effect of a wavefront change between the two images is an overall
constant independent of the actual displacement between the two images;
the variation is only a few micro-pixels. 
If the overall constant is common for both stars, this effect cancels 
for differential astrometry.

\begin{figure}
\begin{center}
\begin{tabular}{c}
\includegraphics[height=6.5cm]{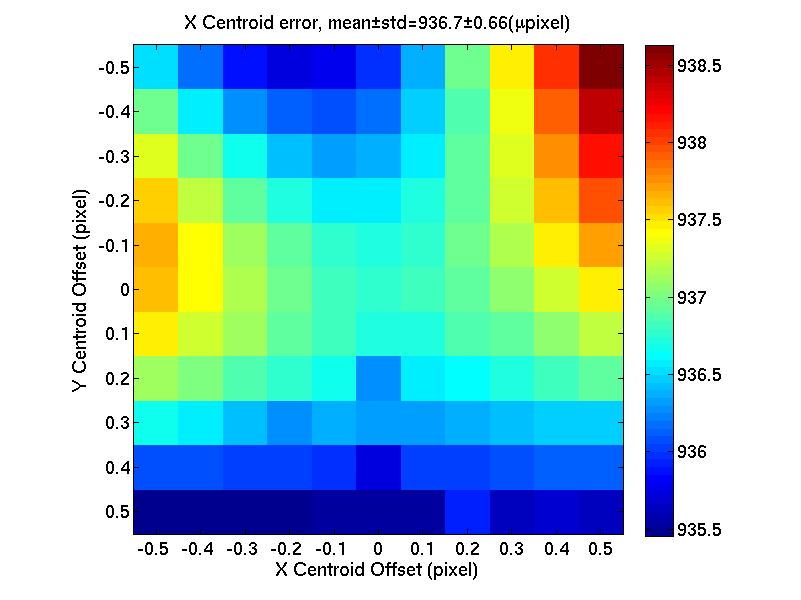}
\includegraphics[height=6.5cm]{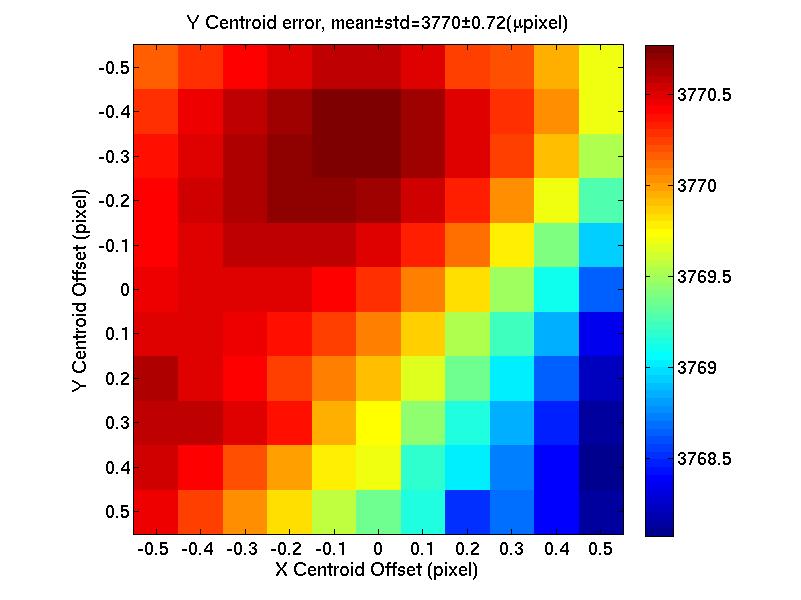}
\end{tabular}
\end{center}
\caption{Centroid estimation errors X (left) and Y (right) in micro-pixel unit for
the second detector. The images are taken with a common low order wavefront error of $\lambda/20$ RMS specified by the first 15 Zernikes polynomials and a differential low order wavefront error of RMS $\lambda/1350$.}
\label{centroidFor2ndCam}
\end{figure}

\begin{figure}
\begin{center}
\begin{tabular}{c}
\includegraphics[height=6.5cm]{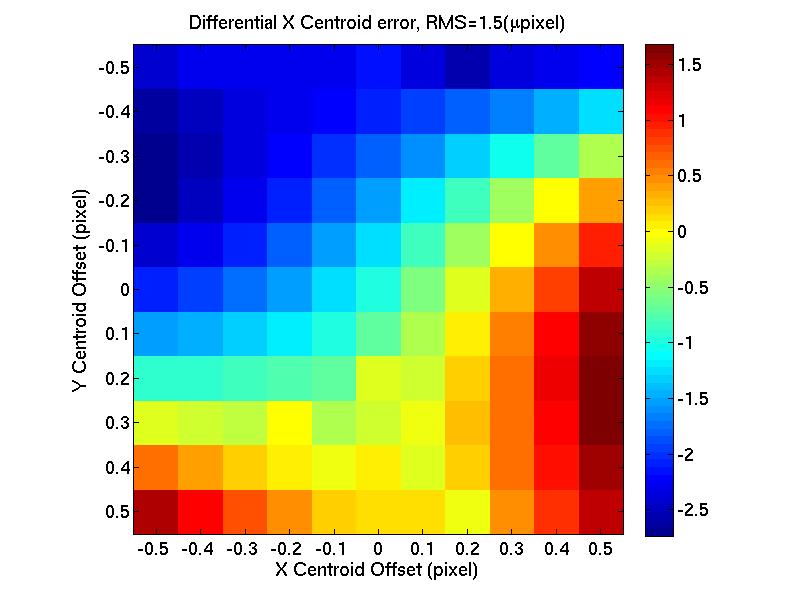}
\includegraphics[height=6.5cm]{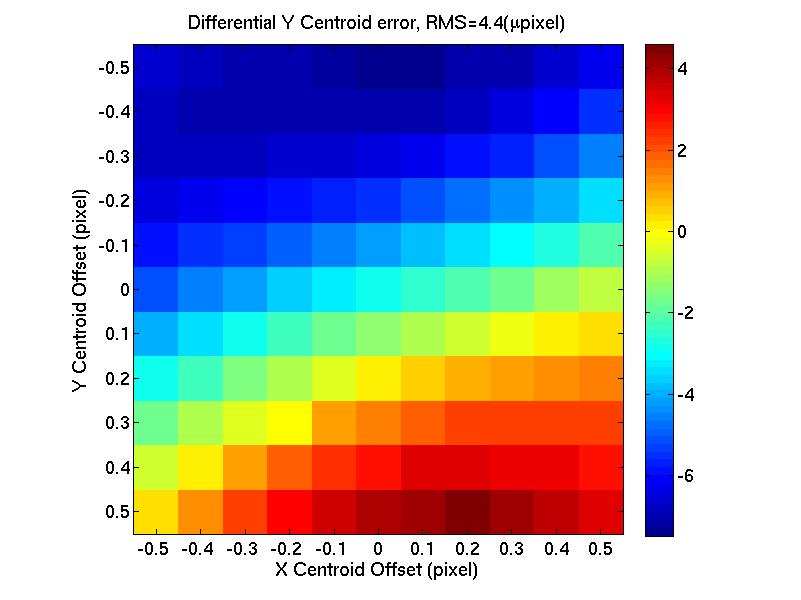}
\end{tabular}
\end{center}
\caption{The difference between the centroid displacement estimation errors
X (left) and Y (right) in micro-pixel unit for two different simulated detectors.
For each detector, the images are taken with a common low order wavefront error of $\lambda/20$ RMS specified by the first 15 Zernikes polynomials and a differential low order wavefront error of RMS $\lambda/1350$.}
\label{diffPixRespWfChangeEffect}
\end{figure}

Because the images of the two stars are at different locations in the
focal plane array, we examine whether the different pixel response functions at
different portions of the detector coupled with the wavefront change leads
to significant error. Fig.~\ref{centroidFor2ndCam} displays the centroid
errors for using the second detector.
To avoid common error cancellation, we put the reference image at [0.25, 0.25]pixel
instead of [0, 0] pixel as for the first detector.
Again, we can see that the dominant effect of wavefront change between the
two images is an overall offset in the centroid estimation. Because the overall
offset is not sensitive to the difference between two detectors,
the overall differential displacement
caused by the wavefront difference cancels leaving a residual of a
few micro-pixel RMS. See Fig.~\ref{diffPixRespWfChangeEffect}.
\newpage

\section{Polychromatic effect}
\label{sec:polychrom}
\begin{figure}
\begin{center}
\begin{tabular}{c}
\includegraphics[height=6.5cm,width=7.5cm]{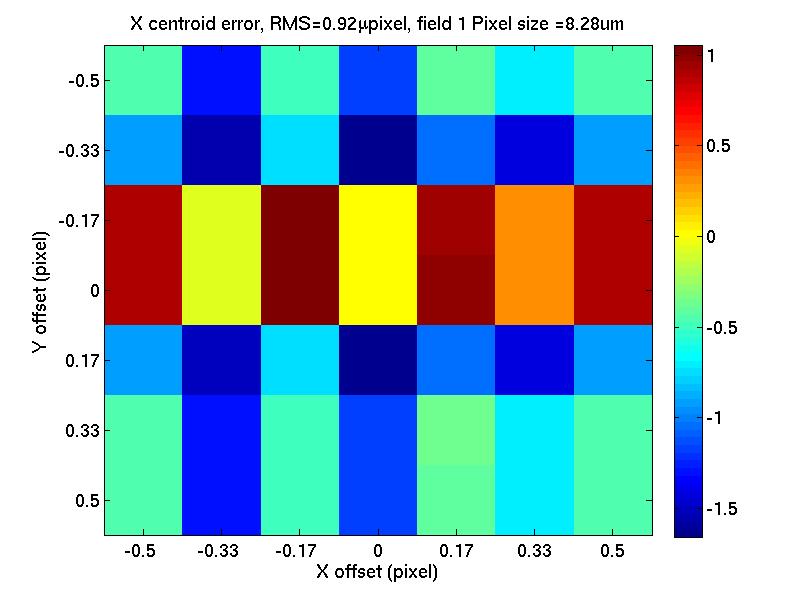}
\includegraphics[height=6.5cm,width=7.5cm]{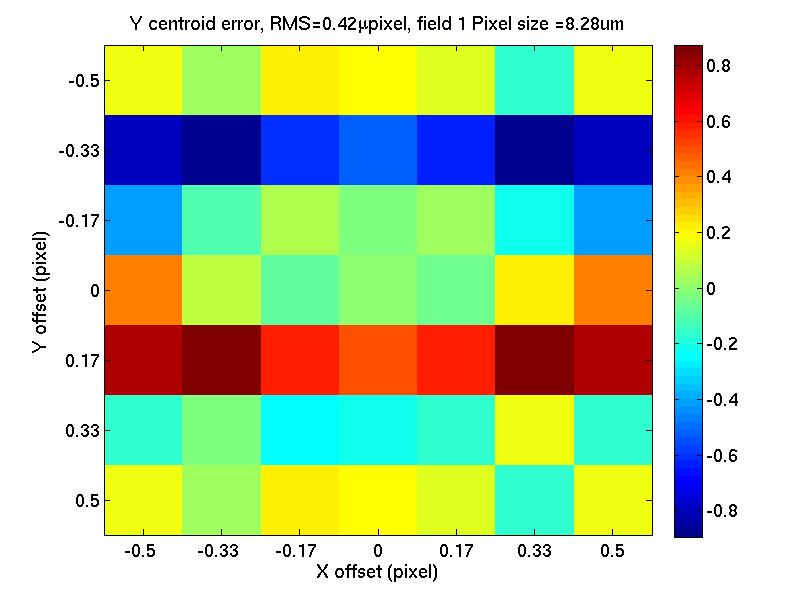}
\end{tabular}
\end{center}
\caption{The estimation errors of the centroid offsets for broadband images.}
\label{polychromatic}
\end{figure}

In this section, we discuss the polychromatic effect.
For white light source like stars, image intensity model~(\ref{pixModelQeFt})
needs an extra integration over the photon wavelengths weighed with 
the source spectrum,
\beq
  I_{mn}(x_c, y_c) =
\infint \! dk_x \!\! \infint\!dk_y \int_0^\infty d\lambda S(\lambda)
  {\cal I} (k_x, k_y, \lambda) {\tilde Q}_{mn}\!(k_x, k_y, \lambda)
  e^{i\left [k_x((m{+}1/2)a{-}x_c){+}k_y((n{+}1/2)a{-}y_c) \right ]}
\label{pixModelQeFtWhiteLight}
\eeq
where $S(\lambda)$ is the source spectral energy density function
and we have included the wavelength dependencies in Fourier transforms
of the monochromatic PSF function ${\cal I}(k_x, k_y, \lambda)$ and
the pixel detection function ${\tilde Q}_{mn}(k_x, k_y, \lambda)$.
As a leading order approximation, we ignore the spectral dependency
in the pixel detection function. The broadband image model has the
same expression as the monochromatic model with the following replacement
\beq
  {\cal I}(k_x, k_y) \to \int_0^\infty d\lambda S(\lambda)
  {\cal I}(k_x, k_y, \lambda) \,.
\eeq
Because the integral over the photon wavelength is a linear operation,
the polychromatic signal is still a bandwidth limited signal, whose
bandwidth is determined by the shortest wavelength.
As far as the pixelated images
are Nyquist sampled, our algorithms works as for the case of monochromatic
source.
Fig.~\ref{polychromatic} displays the centroid offset estimation errors for
simulated chromatic pixelated images pairs displaced by various offsets
along x and y directions assuming an ideal tophat pixel response detection.
We note that the error is sub-micro pixel. 
A slight complication comes from the dependency of pixel response
on the wavelength, which requires pixel response calibration using
metrology at multiple wavelengths.
The results in reference\cite{intraBackIllu} for backside illuminated
CCD shows weak dependency on the wavelengths especially at the longer
wavelength. Because we only need to calibrate the pixel to pixel
variations of the pixel response functions, for differential astrometry,
only the spectral dependency in the inter-pixel variations of the response
functions coupled with the star spectral difference could cause systematic errors. 
We expect this effect to be small in general and can be calibrated
using laser metrology at a few wavelengths. We defer the study of this
to a future work.

\section{Conclusions}
We have presented a systematic frame work for accurate centroid
displacement estimation and detector characterization.
For an ideal detector with all the pixels having the same pixel response function,
the PSF can be accurately reconstructed without any detector calibration.
For realistic detectors whose pixel response functions share a dominant
common portion and small pixel to pixel variations,
it is possible to measure the pixel responses functions in Fourier
space using laser metrology fringes.
Measuring the pixel response in Fourier space is especially convenient for
well sampled images. Keeping a few low order terms (e.g. 3rd
order) in the Taylor series expansion of the Fourier transform in
wave numbers, we can achieve centroid estimations accurate to
a few micro-pixels. This enables micro-arcsecond
level astrometry using telescope images of stars and thus detect earth-like
exo-planets. This method is also applicable to precise photometry.

\section{Acknowledgments}
This work was prepared at the Jet Propulsion Laboratory,
California Institute of Technology,
under a contract with the National Aeronautics and Space Administration.

\end{document}